\documentclass[%
manuscript=article, 
 reprint,
 superscriptaddress,
nofootinbib,
 amsmath,amssymb,
 aps,
prb,
floatfix,
]{revtex4-2}

\usepackage[english]{babel}
\usepackage{graphicx}
\usepackage{dcolumn}
\usepackage{bm}
\usepackage{siunitx}
\usepackage[version=4]{mhchem}
\usepackage{amsmath}
\usepackage{lineno}
\usepackage{url}

\begin{document}

\title{Tailoring Optical Excitation to Control Magnetic Skyrmion Nucleation}

\author{L.-M.\ Kern}
\affiliation{Max Born Institute for Nonlinear Optics and Short Pulse Spectroscopy, 12489 Berlin, Germany}
\author{B.\ Pfau}
\email{bastian.pfau@mbi-berlin.de}
\affiliation{Max Born Institute for Nonlinear Optics and Short Pulse Spectroscopy, 12489 Berlin, Germany}
\author{M.\ Schneider}
\affiliation{Max Born Institute for Nonlinear Optics and Short Pulse Spectroscopy, 12489 Berlin, Germany}
\author{K.\ Gerlinger}
\affiliation{Max Born Institute for Nonlinear Optics and Short Pulse Spectroscopy, 12489 Berlin, Germany}
\author{V.\ Deinhart}
\affiliation{Max Born Institute for Nonlinear Optics and Short Pulse Spectroscopy, 12489 Berlin, Germany}
\author{S.\ Wittrock}
\affiliation{Max Born Institute for Nonlinear Optics and Short Pulse Spectroscopy, 12489 Berlin, Germany}
\author{T.\ Sidiropoulus}
\affiliation{Max Born Institute for Nonlinear Optics and Short Pulse Spectroscopy, 12489 Berlin, Germany}
\author{D.\ Engel}
\affiliation{Max Born Institute for Nonlinear Optics and Short Pulse Spectroscopy, 12489 Berlin, Germany}
\author{I.\ Will}
\affiliation{Max Born Institute for Nonlinear Optics and Short Pulse Spectroscopy, 12489 Berlin, Germany}
\author{C.\ M.\ Günther}
\affiliation{Technische Universität Berlin, Zentraleinrichtung Elektronenmikroskopie (ZELMI), 10623 Berlin, Germany}
\author{K.\ Litzius}
\affiliation{Max Planck Institute for Intelligent Systems, 70569 Stuttgart, Germany}
\author{S.\ Wintz}
\affiliation{Max Planck Institute for Intelligent Systems, 70569 Stuttgart, Germany}
\author{M.\ Weigand}
\affiliation{Max Planck Institute for Intelligent Systems, 70569 Stuttgart, Germany}
\affiliation{Helmholtz-Zentrum Berlin für Materialien und Energie GmbH, 14109 Berlin, Germany}
\author{F.\ Büttner}
\affiliation{Helmholtz-Zentrum Berlin für Materialien und Energie GmbH, 14109 Berlin, Germany}
\author{S.\ Eisebitt}
\affiliation{Max Born Institute for Nonlinear Optics and Short Pulse Spectroscopy, 12489 Berlin, Germany}
\affiliation{Technische Universität Berlin, Institut für Optik und Atomare Physik, 10623 Berlin, Germany}

\begin{abstract}
In ferromagnetic multilayers, a single laser pulse with a fluence above an optical nucleation threshold can create magnetic skyrmions, which are randomly distributed over the area of the laser spot. However, in order to study the dynamics of skyrmions and for their application in future data technology, a controllable localization of the skyrmion nucleation sites is crucial. Here, it is demonstrated that patterned reflective masks behind a thin magnetic film can be designed to locally tailor the optical excitation amplitudes reached, leading to spatially controlled skyrmion nucleation on the nanometer scale. Using x-ray microscopy, the influence of nanopatterned back-side aluminum masks on the optical excitation is studied in two sample geometries with varying layer sequence of substrate and magnetic Co/Pt multilayer. Surprisingly, the masks' effect on suppressing or enhancing skymion nucleation reverses when changing this sequence. Moreover, optical near-field enhancements additionally affect the spatial arrangement of the nucleated skyrmions. Simulations of the spatial modulation of the laser excitation, and the following heat transfer across the interfaces in the two sample geometries are employed to explain these observations. The results demonstrate a reliable approach to add nanometer-scale spatial control to optically induced magnetization processes on ultrafast timescales.
\end{abstract}

\maketitle

\section{\label{sec:level1}Introduction}

In ferromagnetic thin-films with perpendicular magnetic anisotropy (PMA), magnetic skyrmions appear as two-dimensionally localized, solitonic spin textures with spherical topology \cite{fert_2013, tomasello_2015}. Skyrmions are quasiparticles of small size and high stability which are of interest in basic science and potentially for applications due to their emerging topological charge. Femtosecond laser pulses offer an effective way to manipulate magnetization \cite{beaurepaire_1996,barman_2007,malinowski_2008,pfau_ultrafast_2012}, including magnetic skyrmions \cite{je_creation_2018,berruto_laser-induced_2018,novakovic_2020,buettner_2020,Gerlinger_2021}, in a faster and, potentially, more energy efficient fashion than the more established current-induced nucleation mechanism \cite{woo_observation_2016,moreau2016additive,buttner_field-free_2017,legrand2017room,litzius_2017,woo_2018,caretta_2018}. Previous work has already demonstrated that nanometer-scale, room-temperature-stable skyrmions can be nucleated by a single femtosecond laser pulse in Co-based ferromagnetic multilayers \cite{je_creation_2018,buettner_2020,Gerlinger_2021}. To optically nucleate skyrmions, a material-dependent nucleation threshold \cite{buettner_2020,Gerlinger_2021} needs to be overcome first. In a laser fluence regime above this nucleation threshold, the available thermal energy allows the magnetization to reorder after excitation from a high-temperature phase characterized by topological fluctuations into a pure skyrmion state that is typically inaccessible through adiabatic field cycling alone \cite{buettner_2020}.

However, the laser-induced nucleation leads to skyrmions that are spatially distributed in random arrangements within the whole laser-illuminated area where the fluence threshold is exceeded \cite{buettner_2020, Gerlinger_2021}. However, in view of applied skyrmion research, it is important to be able to reproducibly nucleate skyrmions in predefined regions of the sample. In addition, also fundamental studies on skyrmion dynamics based on repetitive pump--probe schemes require deterministic nucleation of skyrmions precisely localized at a specific site. One approach to achieve spatial control over a laser-induced process is to spatially tailor the optical excitation. This has been demonstrated in the past via tight focusing \cite{finazzi_2013}, the use of near-field effects \cite{korff_2014,liu_2015} and the use of nanopatterned reflective masks on the surface of the sample, facing the incident laser beam \cite{korff_2015,weder_2020}. These approaches have in common that they rely on a structural modification of the sample surface, partially inhibiting direct access to the magnetic film. Alternatively, Helium ions were used to modify the magnetic anisotropy landscape to predefine nucleation sites for the skyrmions \cite{kern2022deterministic}. Nonetheless, this method is restricted to materials that can be beneficially modified in this way and the permanent changes in the material may also disturb its dynamical properties. 

Here, we demonstrate a new route to spatially control the optical excitation of a thin film magnetic system with the goal to localize magnetic skyrmion nucleation sites: micrometer to nanometer patterned infrared-reflective masks which are generated on the \emph{back} side of the thin film system, i.e. on the side \emph{not} facing the infrared (IR) laser beam. Note that this approach allows retaining a flat, undisturbed surface of the magnetic thin film system, allowing for free front side access for analytical techniques and in particular for approaches to read out the local magnetization state.

As we will show below, the mechanism for the emergence of a local temperature distribution is based on the infrared reflectivity of the structured aluminum layers, while the thermal conductivity is irrelevant on the subnanosecond timescale on which the skyrmion nucleation occurs. We study the nucleation behavior and spatial arrangement of skyrmions as a function of the geometry of the IR-reflective mask structures by x-ray imaging of the sample magnetization in transmission, enabled by the high transparency of the aluminum structures for soft x-rays. The experimental findings are compared with finite-element simulations allowing us to understand the excitation amplitude in the magnetic layer as a function of the sample layout, in particular including the stacking order of magnetic layer, substrate and reflective mask.

\section{\label{sec:exp}Experimental Details}

\subsection{Direct and Indirect Coupling Geometry}

Magnetic multilayers with a nominal composition of Ta(\SI{3}{nm})/ [Co(\SI{0.6}{nm})/ Pt(\SI{0.8}{nm})]$_{12}$/ Ta(\SI{2}{nm}) and Ta(\SI{3}{nm})/ Pt(\SI{4}{nm})/ [Pt(\SI{2.5}{nm})/Co$_{60}$Fe$_{25}$B$_{15}$(\SI{0.72}{nm})/ MgO(\SI{1.4}{nm})]$_{15}$/ Pt(\SI{2}{nm}) were deposited on \SI{150}{nm} thick, silicon-nitride membranes via DC and RF magnetron sputtering. Throughout the manuscript the multilayers are denoted as Co/Pt and Pt/CoFeB/MgO, respectively.

\begin{figure}
	\begin{center}
		\includegraphics{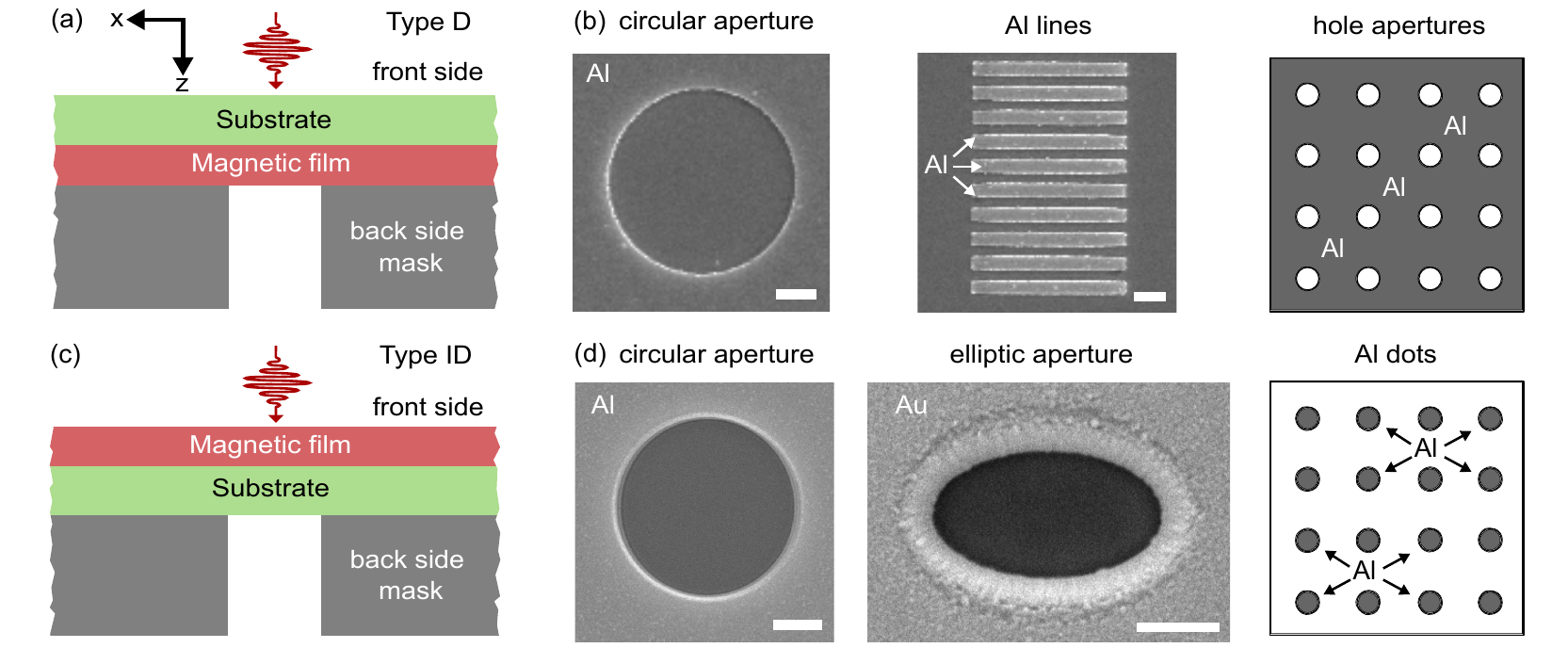}
	\end{center}
	\caption{\label{fig:geometry}\textbf{Scheme of the direct and indirect coupling geometry.} (a) Direct coupling geometry with direct contact of the magnetic film with the back-side reflective mask; and (b) corresponding SEM images of circular apertures, Al lines and a schematic of hole apertures in Al. (c) Indirect coupling geometry with the substrate sandwiched between the magnetic film and the back-side reflective mask; and (d) corresponding SEM images of circular and elliptic apertures and a schematic of Al dots. Scalebars correspond to \SI{1}{\micro m}.}
\end{figure}

Both materials provide an out-of-plane magnetization required for stable skyrmion nucleation \cite{buettner_2020,Gerlinger_2021}. Micro to nano-structured reflective masks were generated on the back side of our multilayer samples. We arranged the stacking order of the substrate, magnetic layer and the mask in two different geometries as presented in Fig.~\ref{fig:geometry}, providing different coupling of the magnetic layer to the back-side reflective mask. While the geometry sketched in Fig.~\ref{fig:geometry}a provides a direct coupling between both layers, the two layers have an indirect coupling in the second geometry as presented in Fig.~\ref{fig:geometry}c, because the interface is mediated by the dielectric substrate. In our experiment, we image the samples with scanning transmission x-ray microscopy (STXM) exploiting the x-ray magnetic circular dichroism (XMCD) providing contrast to the out-of-plane magnetization component. To retain soft-x-ray transparency, the masks are produced from Al films.

In the direct coupling geometry (labeled as \emph{Type D} in the figures), the mask layer was directly prepared on the magnetic film by electron-beam lithography. The mask consists of a \SI{3}{nm} chromium adhesion layer, followed by \SI{200}{nm} aluminum, deposited via electron-beam evaporation. The structures include holes of \SI{2.9}{\micro m} diameter, laminar gratings of \SI{500}{nm} width, separated by \SI{100}{nm} or \SI{500}{nm} spaces and arrays of holes with \SI{150}{nm} diameter and a pitch of \SI{500}{nm}, as presented in Fig.~\ref{fig:geometry}b. Note that the sample was exposed to air between the deposition of the magnetic material and the aluminum mask layer, most likely leading to a thin oxide layer at the interface to the Ta cap layer. In this geometry, the pump laser pulse is incident on the IR-transparent silicon-nitride substrate, prior to reaching the magnetic film.

In the indirect coupling geometry (labeled as \emph{Type ID} in the figures), the mask layer, here a \SI{3}{nm} chro\-mi\-um adhesion layer, followed by \SI{300}{nm} aluminum, was deposited on the silicon-nitride substrate, opposite of the magnetic film. The mask layer was processed with a focused ion beam (FIB) to shape single apertures of \SI{3}{\micro m} diameter and arrays of dots with \SI{150}{nm} diameter, separated by \SI{500}{nm} spacing. In the experiments, the laser pulse impinges directly on the magnetic film, as illustrated in Fig.~\ref{fig:geometry}c.

Experiments using holographic imaging \cite{eisebitt_lensless_2004} were carried out with additional samples in indirect coupling geometry. The mask composition is [Cr(\SI{5}{nm})/Au(\SI{50}{nm})]$_{20}$, and this layer serves the double purpose as holographic optics mask and reflector structure for the IR pulses. The holography mask contains a circular aperture with a diameter of \SI{3.5}{\micro \meter} or elliptic aperture with a long axis of \SI{2.5}{\micro m}. Additional apertures as through-holes (diameter \SI{\sim 40}{nm}) act as sources for providing the reference waves. Note that this mask is opaque to soft x-rays and our field of view is limited to the defining aperture (see Methods section for detailed information on Magnetization Imaging). The IR pulses used to nucleate skyrmions are always incident on the unpatterned magnetic layer, i.e., the back-side reflective mask faces away from the incident IR beam.

\subsection{Magnetization Imaging}

We detected the magnetic textures created via high-resolution x-ray microscopy and x-ray holography. Scanning transmission x-ray microscopy (STXM) was carried out at the MAXYMUS endstation \cite{grafe_visualizing_2019} at the BESSY II electron storage ring, operated by the Helmholtz Zentrum Berlin für Materialien und Energie. Images were recorded in transmission geometry with circularly polarized x-rays, tuned to the Co L$_3$ absorption resonance at a photon energy of \SI{778}{eV} to employ the x-ray magnetic circular dichroism (XMCD) effect for magnetic contrast. The resulting magnetization maps reflect the out-of-plane component of the sample magnetization. We show all STXM images as recorded within a single scan with one particular x-ray helicity. The images, thus, contain both topographic and magnetic contrast, where the topographic contrast is exclusively due to the shape of the mask structures. 
At a photon energy of \SI{778}{eV}, the masks have a nominal transmission \cite{henke_1993,weder_2020} of \SI{88}{\percent} (\SI{83}{\percent}) for the direct (indirect) geometry, keeping the magnetic textures in the Co/Pt multilayers detectable for soft x-rays.

X-ray-holography-based imaging was carried out at the undulator beamline P04 of the syn\-chro\-tron-ra\-dia\-tion facility PETRA III (Hamburg, Germany) in transmission, relying on the same magnetic contrast \cite{eisebitt_lensless_2004}. Details on the setup can be found in Ref.~\cite{buettner_2020}. In contrast to the STXM images, x-ray holography images are reconstructed from a difference of holograms recorded with opposite helicity. Therefore, the images only contain magnetic contrast.

\section{Results}
The general procedure in the experiments was the following: The sample was positioned in an external magnetic field along the $z$-direction, parallel to the magnetization's easy axis and perpendicular to the magnetic film plane. First, we fully saturated the sample in a magnetic field of \SI{200}{mT}. 
We then decreased the applied magnetic field to a value for which the sample remains saturated, but optically induced skyrmion nucleation becomes possible \cite{buettner_2020,Gerlinger_2021}. We imaged the magnetic state of our sample before and after laser excitation and we detect the magnetic skyrmion textures created as a function of the applied laser fluence and the external magnetic field. The range of eligible field values depends on the magnetic material investigated \cite{novakovic_2020,buettner_2020,Gerlinger_2021} and will be presented in further detail in the following for our Co/Pt multilayers. We optically induced skyrmion formation in the magnetic film at the STXM experiment using single IR (\SI{1039}{nm} wavelength) laser pulses of \SI{8}{ps} duration and a focal spot size of \SI{6.5}{\micro m} (full width at half maximum, FWHM) and at the x-ray holography experiment using single IR (\SI{1030}{nm} wavelength) laser pulses of \SI{250}{fs} duration and a focal spot size of \SI{60}{\micro m} (FWHM).

\subsection{\label{subsec:exp_direct}Direct Coupling Geometry}
\begin{figure*}
	\includegraphics{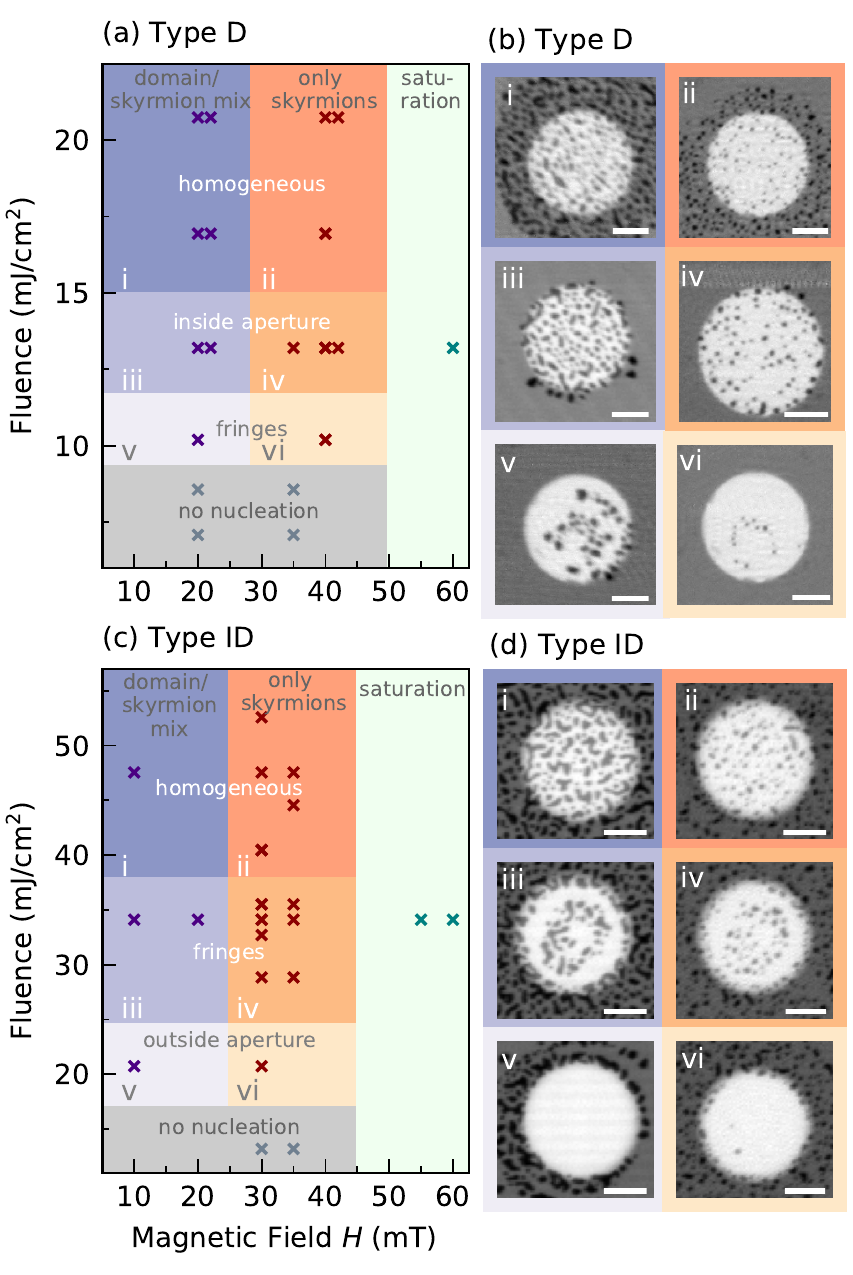}
	\caption{\label{fig:phase_diagrams} \textbf{Geometry-dependent skyrmion configurations in Co/Pt nucleated by a exposure with a single laser pulse.} (a--b) Direct coupling geometry with (a) phase diagram, and (b) representative STXM images for the different regimes identified in (a). (c--d) Indirect coupling geometry with (c) phase diagram, and (d) representative STXM images for the different regimes identified in (c). Scalebars correspond to \SI{1}{\micro m}.}
\end{figure*}

After laser exposure, magnetic textures (black) appear with a net magnetization orientation opposed to that of the previously saturated film (white) (as seen in Fig.~\ref{fig:phase_diagrams}b). In particular, we find small circular domains with sub-\SI{100}{nm} diameter. Previous Lorentz transmission electron microscopy analysis of a very similar material proved that these domains are skyrmions \cite{buettner_2020}. The back-side reflective mask present in the field of view appears as a region with dark-gray background (higher absorption). In contrast, the mask-free aperture results in a light-gray background (lower absorption).

We observe different skyrmion patterns in distinct regimes of the applied laser fluence and applied field. Figure~\ref{fig:phase_diagrams}a summarizes these experimental findings in a phase diagram of magnetic textures observed. Exemplary STXM images illustrating the individual phases are presented in Fig.~\ref{fig:phase_diagrams}b as indicated by color coding.

Tuning the incident laser fluence allows us to affect if and where skyrmions nucleate (Fig.~\ref{fig:phase_diagrams}a,b). We observe the existence of a sharp fluence threshold around \SI{9.5}{mJ/cm^2}, below which skyrmion nucleation is impossible in any of the field regimes (saturated images not shown). This threshold behavior is known from previous experiments without any reflective mask \cite{buettner_2020,Gerlinger_2021}. In a narrow laser fluence window from \SI{9.5}{mJ/cm^2} to \SI{12}{mJ/cm^2}, pearl-on-a-chain-like skyrmion patterns appear in the aperture region, illustrated in Fig.~\ref{fig:phase_diagrams}b, panels v and vi. Skyrmion formation in the masked region is still inhibited. As depicted in panels iii and iv, skyrmions exclusively nucleate in the region without mask in a fluence regime from \SI{12}{mJ/cm^2} to \SI{15}{mJ/cm^2}, homogeneously covering the entire area of the aperture. Further increasing the fluence leads to the nucleation of skyrmions also in the masked region, shown in panels i and ii.

Varying the external magnetic field reveals three regimes: from \SI{5}{mT} to \SI{28}{mT}, we observe a mix of stripe domains and skyrmions.  Between \SI{28}{mT} and \SI{50}{mT}, we nucleate only skyrmions. A field above \SI{50}{mT} prevents the stable generation of laser-induced skyrmions at a fluence of \SI{13.5}{mJ/cm\squared}, and the magnetic film remains saturated. We did not explore if skyrmions can be formed at higher fluence. We note that the magnitude of the applied field does influence the skyrmion density, investigated in detail in Ref.~\cite{Gerlinger_2021}. 

To summarize our main observation for the direct coupling geometry, we find that for a particular laser fluence range the Al back-side reflective mask inhibits skyrmion nucleation, while at the same time skyrmion nucleation remains possible in the area without mask. 

\subsection{\label{subsec:geo_indirect}Indirect Coupling Geometry}
In Fig.~\ref{fig:phase_diagrams}c, we present the phase diagram for the indirect coupling geometry. The individual fluence and field regimes that support skyrmion nucleation are shown in the STXM images in Fig.~\ref{fig:phase_diagrams}d.

We again find the general fluence threshold for skyrmion nucleation, below which skyrmions do not nucleate, irrespective of the applied field (saturated images not shown). However, this threshold of around \SI{17}{mJ/cm^2} is now higher than in the direct coupling geometry.
In the fluence range from \SI{17}{mJ/cm^2} to \SI{25}{mJ/cm^2}, skyrmions only nucleate in the area covered by the reflective mask on the back and mainly not inside the open aperture region, shown in Fig.~\ref{fig:phase_diagrams}e, panels v and vi. Increasing the laser fluence into the window from \SI{25}{mJ/cm^2} to \SI{38}{mJ/cm^2} leads to skyrmion nucleation in the mask region and, in addition, to the formation of prominent skyrmion ring patterns inside the open aperture, presented in panel iii and iv. The ring patterns will be discussed in more detail below. Above this fluence window, skyrmion nucleation is allowed in both, masked and unmasked, regions, as shown in panels i and ii.

The kind of of magnetic textures nucleated depends on the applied magnetic field in a similar way as for the direct coupling geometry: up to \SI{25}{mT}, we obtain a mix of stripe domains and skyrmions. Increasing the magnetic field leads to the regime where only skymions nucleate (\SI{25}{mT} to \SI{45}{mT}). For magnetic field values above \SI{45}{mT}, the sample remains saturated after laser irradiation. Again, we did not aim to determine the field threshold precisely, we only show that for high enough fields, the sample remains in saturation.

To summarize our main observations for the indirect coupling geometry, we identify that for a particular fluence range skyrmion nucleation is only allowed in the area with back-side reflective mask, while it is inhibited in open regions without mask. This behavior is inverse to our observations for the direct coupling geometry. Note that via the two different coupling geometries, it is possible to laser-generate skyrmions either almost exclusively inside the aperture region (Fig.~\ref{fig:phase_diagrams}b, iv) or almost exclusively outside the aperture region (Fig.~\ref{fig:phase_diagrams}d, vi).

\subsection{Nanometer-Scale Localization}
So far, the back-side reflective mask layer and the resulting skyrmion localization were structured on a micrometer-scale. In the following, we present results for localization on the nanometer-scale in order to obtain a position control approaching the size of individual skyrmions.

\begin{figure}
	\includegraphics{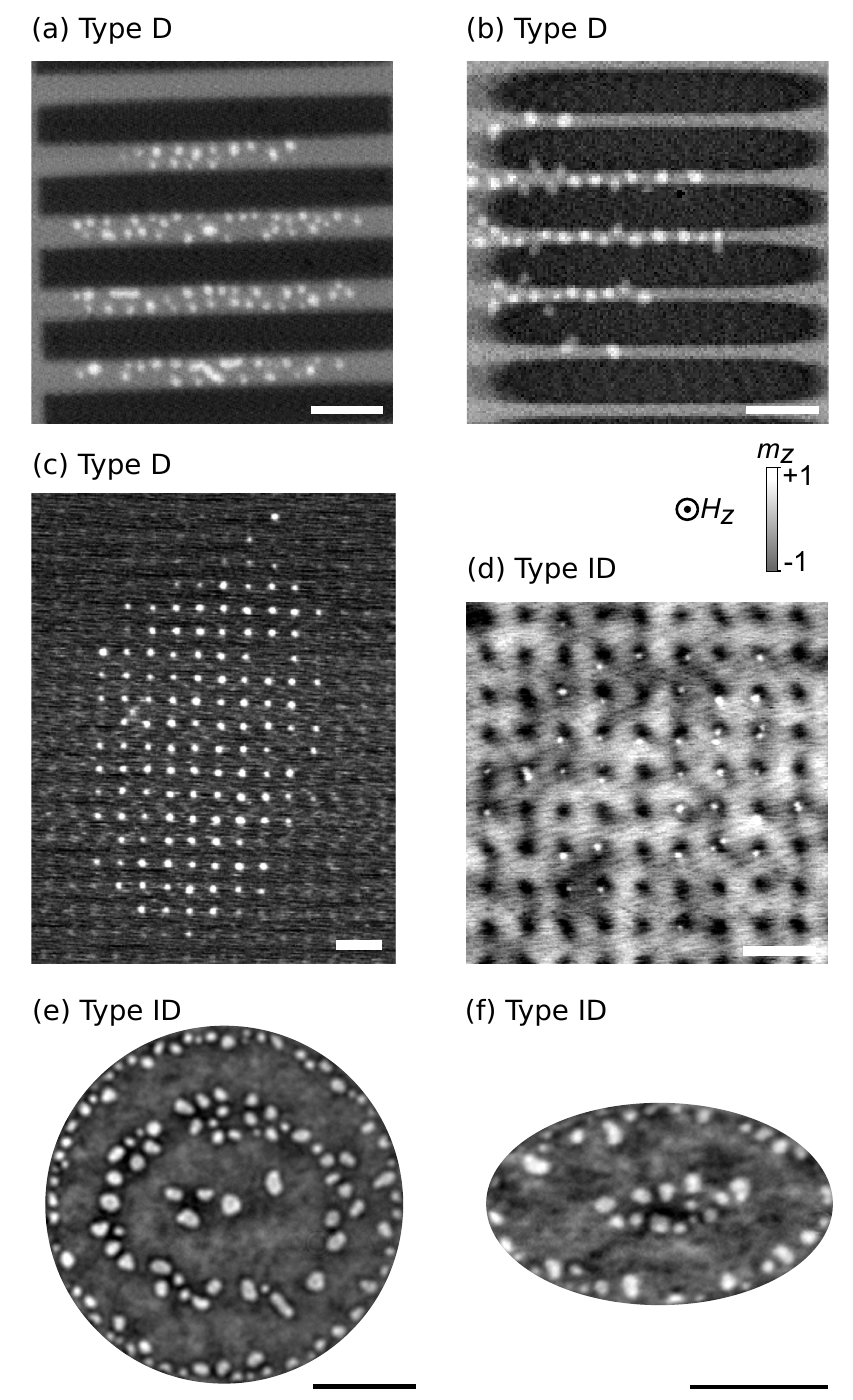}
	\caption{\label{fig:data}\textbf{Localized optical skyrmion nucleation at the nanometer scale.} (a--c) Direct coupling: optical nucleation in laminar gratings with variable spaces ((a) \SI{500}{nm}, (b) \SI{100}{nm}), and (c) array of holes with \SI{150}{nm} diameter. Skyrmions (white dots) form in regions without Al-mask (array of gray holes). (d--f) Indirect coupling: (d) optical nucleation in array of dots with \SI{150}{nm} diameter. Skyrmions (white dots) form in regions with Al-mask (array of black dots). Additional near-field effects cause (e) circular or (f) elliptic skyrmion rings in regions without mask. Scalebars correspond to \SI{1}{\micro m}.}
\end{figure}

First, we investigate mask patterns arranged in a line grating (Fig.~\ref{fig:data}a,b) which were fabricated in the direct coupling geometry. The measurement in Fig.~\ref{fig:data}a was performed in the Co/Pt multilayer, the measurement in Fig.~\ref{fig:data}b in the Pt/CoFeB/MgO multilayer. The Al bars again appear as dark areas. Tuning the laser fluence to the appropriate regime between \SI{15}{mJ/cm}$^2$ (Fig.~\ref{fig:data}a, \SI{40}{mT}) and \SI{25}{mJ/cm}$^2$ (Fig.~\ref{fig:data}b, \SI{35}{mT}), we find that skyrmions almost exclusively nucleate in the gap between the bars, i.e. in a region confined to a gap of \SI{500}{nm} (a) or \SI{100}{nm} (b), respectively. For the \SI{100}{nm} gap, this results in a single line arrangement of the skyrmions within each gap region. Here, skyrmions appear as white dots.

As an example of localization in two dimensions, we present a hole array of \SI{150}{nm} holes, separated by \SI{500}{nm}, in the direct coupling geometry (Fig.~\ref{fig:data}c, fluence: \SI{25}{mJ \per cm \squared}, \SI{30}{mT}) and a dot array of \SI{150}{nm} dots, separated by \SI{500}{nm}, in the indirect coupling geometry (Fig.~\ref{fig:data}d, fluence: \SI{17}{mJ \per cm \squared}, \SI{45}{mT}). In both cases, we achieve localization of single skyrmions at the nanometer scale. Moreover, note that we can optimize for skyrmion nucleation either in the masked or unmasked regions such that we can employ opposite layouts (holes vs.\ dots) to control skyrmion nucleation, demonstrating the versatility of the structured back-side reflective-mask approach.

In Section~\ref{subsec:geo_indirect}, we reported on prominent ring-shaped formations of skyrmions inside the mask-free circular aperture appearing in a narrow regime of the laser fluence. To investigate the nature of these features in more detail, magnetization images, presented in Fig.~\ref{fig:data}e,f, were recorded with x-ray holography on a Co/Pt multilayer sample in the indirect coupling geometry using the same procedure as before. At a fluence of \SI{13}{mJ \per cm \squared} and an applied field of \SI{62}{mT}, we observe that the skyrmions arrange in prominent ring shapes. At lower fluences (\SI{12}{mJ \per cm \squared}) the FOV remains empty after excitation, at higher fluences (\SI{16}{mJ \per cm \squared}) the FOV is homogenously filled with skyrmions (images not shown). Comparing circular and elliptical back-side reflective mask apertures, we observe that the shape of the rings closely follows the aperture shape. Data on differently sized apertures (not shown) suggests that the distance between the staggered rings is approximately constant and independent of the aperture's size and shape. With \SI{\sim 500}{nm} distance these rings suggest near-field effects dependent on the IR wavelength, which is about twice the separation between skyrmion rings.

Here, we have realized localized nucleation on the nanometer-scale via an optical excitation of the magnetic film in two different geometries varying the layer sequence of substrate and magnetic film to either generate skyrmions in a line (direct coupling), in holes (direct coupling) or on dots (indirect coupling) of our reflective mask. In addition, we find evidence for the presence of near-field-effects in a particular fluence range, which could be further exploited in the future. In the following, we complement our experimental findings with simulations of the time and space-dependent temperature distribution in the sample upon laser excitation, in order to understand the mechanisms at work in the different coupling geometries.

\section{Simulation of the Laser-induced Absorption and Heat Transfer}

\subsection{Implementation}
We carried out simulations of the electric-field distribution in the frequency domain to model our optical IR excitation and the resulting absorption in the magnetic film. The laser pulse excitation is normally incident on the sample surface and represented by a plane wave with a Gaussian pulse distribution in the time domain, centered around \SI{0}{ps}, with a wavelength of \SI{1030}{nm}, a pulse duration of \SI{250}{fs} FWHM and an incident fluence of \SI{15}{mJ/cm^2}. The initial temperature is set to \SI{293.15}{K}.  The simulation area is terminated by a perfect electric conductor boundary condition in the lateral dimension and by a perfectly matched layer (i.e., approximating an extended vacuum) in the $z$-direction, suppressing reflections of the laser at the boundaries of the simulation area.

We combined the simulation of the electromagnetic wave with a subsequent heat transfer across the interfaces, based on the time-dependent, two-dimensional two-temperature model described by a simplified system of partial differential equations for the electronic and lattice heat-bath temperatures, as presented in \cite{weder_2020}. The simulation area is thermally insulated, except for the aluminum film that is coupled to a heat bath at room temperature.

The equations are numerically solved using the finite-element method (FEM) and a commercial-grade backward differentiation formula solver [COMSOL multiphysics package, COMSOL AB, Stockholm]. We implemented a two-dimensional model, representing a cross section of our sample in cartesian coordinates $x$ and $z$. The sample simulated comprises a \SI{150}{nm} thick silicon-nitride substrate, the magnetic layer and a \SI{350}{nm} thick aluminum back-side mask layer. Additional simulations confirmed that the simulations of the heat transfer result in temperature deviations of only less than \SI{5}{\percent} for an aluminum layer thickness variation between \SI{25}{nm} and \SI{500}{nm}. The full magnetic material stack is approximated in an effective-medium approach with averaged material parameters according to the composition of the magnetic film. All material parameters used are listed in Tab.~\ref{tab:table1}.

The simulated sample has a total width of \SI{20}{\micro m}, with an aperture in the Al mask of \SI{3}{\micro m} diameter located in the center, approximately matching our experimental geometry of the results presented in Fig.~\ref{fig:phase_diagrams}. However, note that the total width of \SI{20}{\micro m} is a generic choice assuring the computational feasibility as well as a dimension substantially large enough to simulate an electric field distribution in the lateral dimension as well as in-plane heat dissipation on the length-scale of the aperture. The equations are solved on mesh nodes with edge lengths ranging from \SI{1}{nm} to \SI{70}{nm} depending on the local structure size. 

\begin{table*}
	\caption{\label{tab:table1}
		Element-specific parameters used in the FEM simulations of the sample temperature evolution after laser excitation.}
	\begin{tabular}{ccccccc}
			&$n$ &$k$ &$c_{e}$ (\SI{}{Jm^{-3}K^{-1}}) & $c_{l}$ (\SI{}{Jm^{-3}K^{-1}}) &$d_{l}$ (\SI{}{Wm^{-1}K^{-1}}) &$g_{el}$ (\SI{}{Wm^{-3}K^{-1}}) \\
			\hline
			Al& \num{1.07} \cite{mcpeak} & \num{8.87} \cite{mcpeak} & $c_{e}\,^\dagger$ \cite{lin_2008} & \num{2.42}$\times$\num{e6} \cite{alu}
			& \num{237} \cite{alu} & $g_{el}\,^\dagger$ \cite{lin_2008}  \\
			Al$_2$O$_3$& \num{1.75} \cite{querry:1985} & \num{0} \cite{querry:1985} & \num{0} & \num{3.52}$\times$\num{e6} \cite{accuratus}
			& \num{25.08} \cite{shackelford_crc_2001} & \num{0} \\
			Au& \num{0.16} \cite{olmon_optical_2012} & \num{6.85} \cite{olmon_optical_2012} & $c_{e}\,^\dagger$ \cite{lin_2008} & \num{1.74}$\times$\num{e6} \cite{gray:1972} & \num{318} \cite{gray:1972} & $g_{el}\,^\dagger$ \cite{lin_2008} \\
			Co$^\ast$& \num{2.82} \cite{johnson:1974} & \num{5.67} \cite{johnson:1974} & \num{704}$T_e$ \cite{gray:1972} & \num{3.73}$\times$\num{e6} \cite{gray:1972} & \num{100} \cite{gray:1972} & \num{9.3}$\times$\num{e17} \cite{hohlfeld:2000} \\
			Pt& \num{3.53}\cite{rakic:1998} & \num{5.84} \cite{rakic:1998} & \num{740}$T_e$ \cite{gray:1972} & \num{2.78}$\times$\num{e6} \cite{gray:1972} & \num{72} \cite{gray:1972} & \num{2.5}$\times$\num{e17} \cite{hohlfeld:2000} \\
			Si$_3$N$_4$& \num{2.01} \cite{luke_broadband_2015} & \num{0} \cite{luke_broadband_2015} & \num{0} & \num{2.54}$\times$\num{e6} \cite{para_sin} & \num{30} \cite{para_sin} & 0 \\
			Ta$^\#$& \num{0.97} \cite{ordal:1988} & \num{5.09} \cite{ordal:1988} & \num{740}$T_e$ \cite{gray:1972} & \num{2.78}$\times$\num{e6} \cite{gray:1972} & \num{72} \cite{gray:1972} & \num{2.5}$\times$\num{e17} \cite{hohlfeld:2000} \\
			\hline
			\multicolumn{7}{l}{$^\dagger$We use tabulated temperature-dependent values for aluminum and gold.}\\
			\multicolumn{7}{l}{$^\ast$For the simulation of the effective magnetic film, the thermal and coupling parameters $c_{e}$, $c_{l}$, $d_{l}$,}\\ 
			\multicolumn{7}{l}{and $g_{el}$ are averaged according to the composition of the magnetic multilayer.}\\
			\multicolumn{7}{l}{$^\#$The thin Ta layer is simulated using the parameters of Pt.}
	\end{tabular}
\end{table*}

\subsection{Geometry-dependent absorption and temperature depth profiles}

\begin{figure}
	\includegraphics{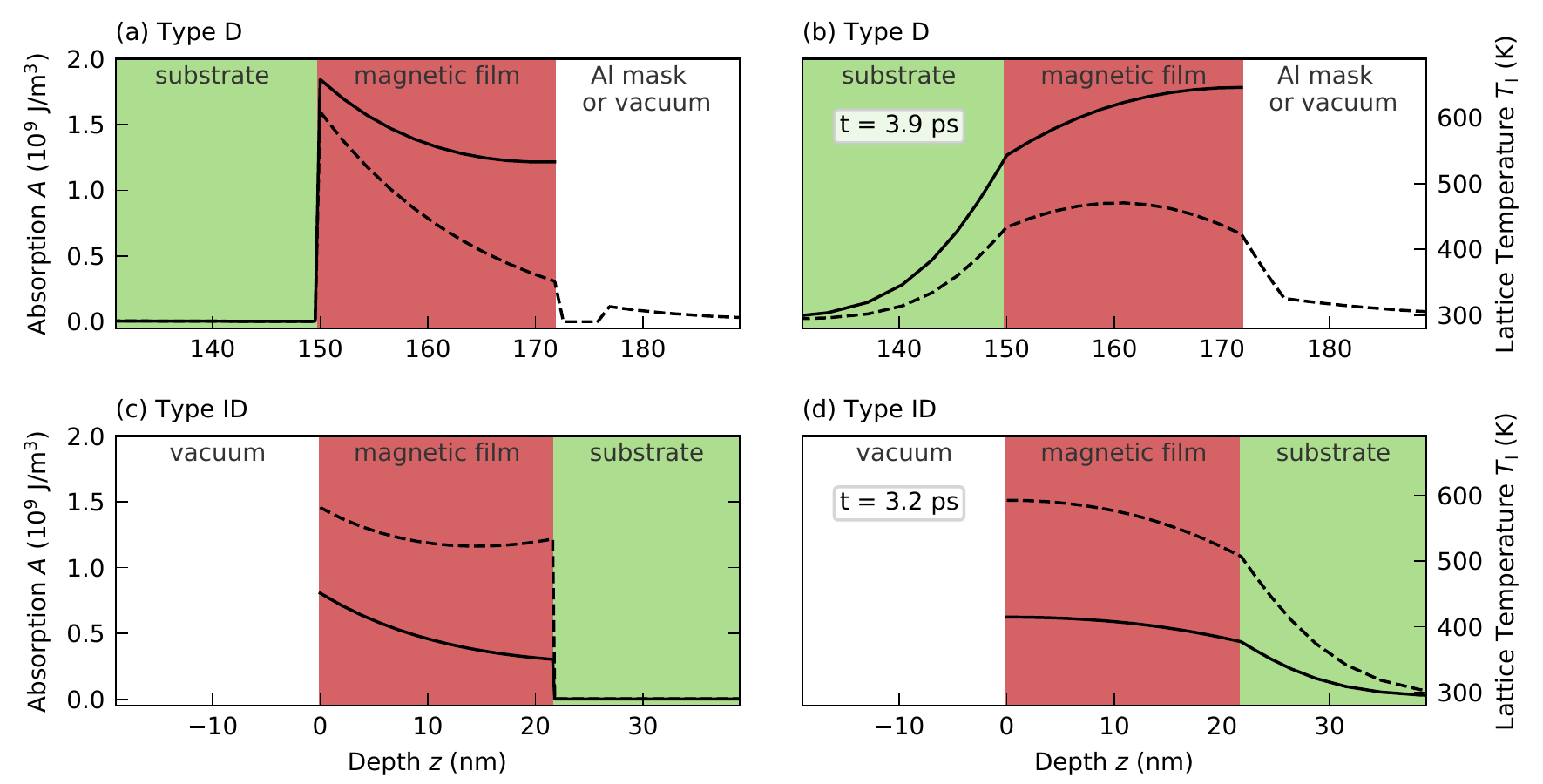}
	\caption{\label{fig:abs}\textbf{Simulated depth dependence of IR absorption and lattice temperature profiles.} Solid curves reflect a central position in the magnetic film without mask, i.e., above the open aperture within the back-side reflective mask, while dashed curves denote a central position in the magnetic film under the mask. (a) Depth profile of absorption and (b) maximum lattice temperature for direct coupling at $t=\SI{3.9}{ps}$. Note that an additional \SI{4}{nm} oxide layer between magnetic film and aluminum mask is added in case of the direct coupling geometry to account for exposing the sample to air during the transfer between DC magnetron sputtering and electron-beam evaporation in the sample's fabrication process. (c) Depth profile of absorption and (d) maximum lattice temperature for indirect coupling at $t=\SI{3.2}{ps}$. Note that the maximum lattice temperatures are reached at the opposite interfaces of the magnetic film for the two situations with and without Al mask.}
\end{figure}

We performed complementary simulations of the laser-induced excitation, by modelling the electric field amplitude, the absorption, and the resulting heat transfer across the interfaces in both geometries (see Methods section for detailed information).
In Fig.~\ref{fig:abs}, the absorption of the laser pulse is shown as a function of the position in the material along the depth direction ($z$-axis) for both geometries.

We make two important observations: First, for the same laser fluence, the absorbed power in the multilayer is generally higher for the direct coupling geometry. Due to its high transparency, the substrate acts as an index-matching layer. This distributes the refractive index change for the incident IR radiation over two interfaces and consequently reduces the overall reflectivity (to \SI{\sim 74}{\percent} in regions with Al mask and to \SI{\sim 61}{\percent} in regions without mask). In contrast, the laser is immediately incident on the magnetic layer in the indirect coupling geometry and \SI{\sim 85}{\percent} are reflected off the surface in regions with Al mask, and \SI{\sim 95}{\percent} are reflected in regions without mask. Second, the presence of the back-side reflective masks affects the absorbed power in regions with and without mask differently in the two geometries.

In the direct coupling geometry, shown in Fig.~\ref{fig:abs}a, more energy is absorbed in regions of the magnetic film \emph{without} reflective mask as compared to the regions \emph{with} reflective mask. In the regions with Al mask, we observe almost no back-side reflection at the metal/metal interface due to the similar refractive indices, and thus an instant decrease of the absorption profile in the magnetic film. In regions without mask, the absorbed energy in the magnetic film increases due to back-reflections from the metal/vacuum interface at the sample's back side. 

In the indirect coupling geometry, shown in Fig.~\ref{fig:abs}c, more energy is absorbed in regions of the magnetic film \emph{with} back-side reflective mask as compared to the unmasked regions. The incoming beam is reflected at every interface in the material stack, and the back-reflection at the substrate/aluminum interface (not shown in the plot) is heating up the magnetic layer again. At positions \emph{without} Al mask, however, the substrate serves as an index matching layer at the exit of the sample to the vacuum reducing reflections. The regions without back-side reflective mask therefore absorb less energy.

In both cases, we present the corresponding depth profiles for the resulting lattice temperature in Fig.~\ref{fig:abs}b,d. The maximum lattice temperature is reached at slightly different times after the laser excitation in both layouts and in regions with or without mask: direct coupling: \SI{3.9}{ps} for regions without mask, \SI{2.8}{ps} for regions with mask; indirect coupling: \SI{3.2}{ps} for regions without mask, \SI{4}{ps} for regions with mask. Consistent with our observations on the IR absorption, the region without back-side reflective mask reaches a higher lattice temperature in the direct coupling geometry, whereas the region with mask takes higher temperature values in the indirect coupling geometry. The calculated profiles are in good agreement with our experimental observations in Fig.~\ref{fig:phase_diagrams}, revealing an inverted excitation behavior when comparing both sample geometries.

\subsection{Lateral temperature profile}
\begin{figure}
	\includegraphics{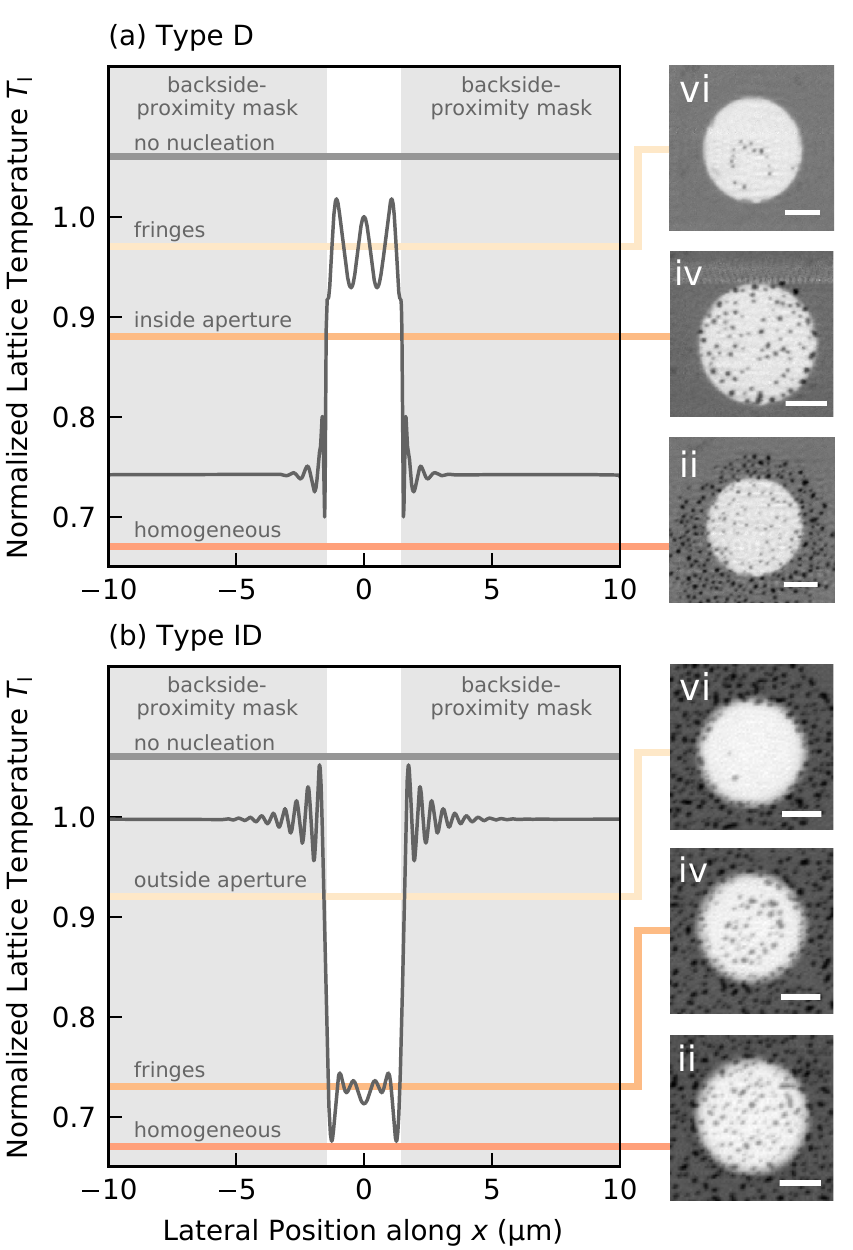}
	\caption{\label{fig:sim_profile}\textbf{Simulated lateral temperature profile.} Lateral temperature profile in the center of the magnetic film for (a) direct coupling and (b) indirect coupling geometry. The orange-shaded horizontal lines in (a) and (b) represent relative, qualitative estimates of the material-dependent nucleation threshold that has been overcome in the corresponding regimes in the phase diagram in Fig.~\ref{fig:phase_diagrams}a and c, respectively, and which is supported by experimental observations as indicated. Scalebars correspond to \SI{1}{\micro m}. If the magnetic film remains saturated after laser excitation, the nucleation threshold has not been overcome, denoted as the regime 'no nucleation'.}
\end{figure}

We now take a closer look at the spatial temperature distribution in the magnetic film plane. In Fig.~\ref{fig:sim_profile}a,b, we present the simulated lattice temperature profile in the film plane (along the $x$-axis) at \SI{3.9}{ps} (direct coupling geometry) and \SI{3.2}{ps} (indirect coupling geometry) after laser excitation, normalized to the maximum temperature. At these times, the region without mask reaches its maximum lattice temperature in the respective coupling geometry. The normalized temperature profile is almost independent of the laser fluence at these early times after excitation and the absolute temperatures reached scale linearly with the laser fluence. Displaying the normalized temperatures allows us to draw horizontal lines, qualitatively indicating the relative position of the nucleation threshold temperature to the sample temperature for the different nucleation regimes presented in Fig.~\ref{fig:phase_diagrams}a,c at different laser fluences. In particular, note that for illustration purposes we have shifted the position of the threshold temperature rather than the sample's temperature for the different regimes. By selecting a particular threshold line for one of the regimes, one can directly read off the sample's nucleation behavior: If the sample temperature is above the threshold, skyrmions will nucleate while nucleation is suppressed below the threshold line. This dependence directly leads to the skyrmion nucleation patterns observed in Fig.~\ref{fig:phase_diagrams}b,d. 

On top of the temperature variation caused by the reflective mask patterning, we observe spatial temperature oscillations of approximately \SI{\pm 20}{K} at early times until \SI{100}{ps}, resulting from near-field interference effects of the laser beam inside the aperture in both geometries. We directly associate the ring like structures in the skyrmion formation observed in Fig.~\ref{fig:phase_diagrams}b,d and Fig.~\ref{fig:data}e,f with these temperature modulations as indicated by the respective threshold line in Fig.~\ref{fig:sim_profile} crossing the temperature modulations.

\section{Discussion}
With our approach of tailored optical excitation, we demonstrated the optical generation of magnetic skyrmions in a thin magnetic film with a position control of about \SI{100}{nm} in one or both lateral dimensions. This is achieved via the use of Al masks below the magnetic film, i.e., on the side opposite to the incident laser pulse. We compare two geometries with different interface coupling of the Al layer to the magnetic multilayer: with and without \SI{150}{nm} silicon-nitride spacer. Interestingly, we find that for a suitable choice of laser excitation and magnetic field, these two geometries lead to an ``inverted'' skyrmion nucleation behavior, allowing us to selectively generate skyrmions either inside or outside of the masked region.

While one may initially expect that the Al layer acts as a local heat sink via direct or indirect thermal coupling of the magnetic film to the Al, we find that this is not the relevant mechanism. Simulations of the depth-dependent absorption of the IR laser pulse for the different geometries allow us to predict the IR absorption and the resulting lattice temperature evolution in the magnetic film, based on a two-temperature model. We clearly see that it is the modification of the initial IR absorption and reflection which lead to significantly different temperature profiles for the two coupling situations. The mechanism is the same in both geometries: At intermediate laser fluences, skyrmion nucleation only occurs in regions where the temperature directly after excitation is high enough to overcome the single-pulse nucleation threshold which was shown to be very sharp \cite{buettner_2020,Gerlinger_2021}. In the direct coupling geometry, this leads to skyrmion nucleation in regions without back-side reflective mask at lower fluences compared to masked regions. In the indirect geometry, on the other hand, nucleation in masked areas is promoted. We demonstrated that our approach can be employed to localize nanometer-scale skyrmions inside 2D reservoir-like areas \cite{zazvorka2019thermal}, 1D chains of skyrmions and bit-patterns of single skyrmions. 

Based on this mechanism, the simulations also predict the fringe-like arrangements of the skyrmions which we find in the experiments in a narrow fluence window. In the indirect geometry in Fig.~\ref{fig:phase_diagrams}d, regimes iii and iv, these formations emerge inside the unmasked aperture while the region under the mask already shows randomly distributed skyrmions throughout the magnetic film. In Fig.~\ref{fig:data}e,f we show high-resolution images of these fringes. A comparison of the experimental results with the simulated spatial distributions of the temperature strongly suggests that the fringes originate from the fine-structure in the temperature in particular within the aperture in the Al mask. In turn, these short-range modulations are due to wavelength dependent, near-field interference effects shaping the lateral laser intensity distribution in the magnetic layer \cite{korff_2014}.
Our results show that transient modulations in temperature distribution occurring only picoseconds after the excitation and on nanometer lengthscale in the magnetic film are decisive in determining skyrmion nucleation sites when they only locally lead to a crossing of the nucleation threshold. 
We note that such fringes are unexpected if the Al layer's function as a heat sink would predominately affect the localization, because the nanometer-scale modulations vanish within tens of picoseconds after excitation due to lateral heat diffusion before the magnetic layer cools down via the Al layer. Beyond the localization of skyrmion positions provided by the micro- and nanopatterning of the Al reflector layer, the existence of these near-field effects provides an additional handle to further tailor the localized optical generation of skyrmions via suitable reflector mask designs.

\section{Conclusion}
Via nanometer-resolution magnetization imaging, we have investigated how the photo-induced nucleation of magnetic textures with reverse ma\-gne\-ti\-za\-tion---skyrmions in our samp\-le sys\-tems---can be localized with regard to their lateral position in a thin magnetic film. We demonstrated an approach based on Al masks on the back side of the sample, acting as IR reflectors of the incident laser beam. By changing the sequence in the layer structure, we could determine whether the Al layer acts as a ``positive'' or ``negative'' mask in the sense that it either increases or decreases the transient peak temperature in the magnetic layer. We achieved a precise localization of skyrmion nucleation of about \SI{100}{nm} in \num{1}D and \num{2}D, corresponding to about one skyrmion diameter in our materials. Furthermore, we observed the influence of near-field interference effects, allowing for an additional means of lateral spatial control which can be exploited by an optimized mask design in the future. We note that the detectable presence of near-field effects implies that it may be possible to take advantage of plasmonic effects as well. Simulations of the laser absorption and temperature evolution in the different layers for both geometries are in good agreement with our experimental observations. These simulations substantiate a mechanism where the Al mask spatially structures the optical excitation, rather than acting as a heat sink. Finally, we would like to stress that the optical localization approach demonstrated here does not require any modifications at the surface of the sample, enabling mechanically and optically unobstructed access to the magnetic thin film surface for, e.g., local magnetic reading. Our back-side mask approach can directly be applied also for the localization of other types of photo-induced switching or modification processes governed by a threshold behavior, such as all-optical switching of ferrimagnetic materials \cite{mangin2014engineered,steinbach2022accelerating}.

\section{Acknowledgements}
Measurements were carried out at BESSY II (Helmholtz-Zentrum Berlin, HZB) and at PETRA III (DESY). We thank Helmholtz-Zentrum Berlin for the allocation of synchrotron-radiation beamtime. We acknowledge DESY (Hamburg, Germany), a member of the Helmholtz Association HGF, for the provision of experimental facilities at PETRA III, beamline P04. Financial support from the Leibniz Association via Grant No. K162/2018 (OptiSPIN) and the Helmholtz Young Investigator Group Program is acknowledged.


%

\clearpage

\end{document}